\documentclass[12pt]{article}
\usepackage{axodraw,mathrsfs}
\makeatletter
\@addtoreset{equation}{section}

\makeatother
\title{\centerline{\normalsize  SINP/TNP/06-12 \hfill hep-ph/0604009}
{\bf Soft breaking of $L_\mu-L_\tau$ symmetry: Light neutrino
  spectrum and Leptogenesis}}
\author{{\bf Biswajit Adhikary\footnote{biswajit.adhikary@saha.ac.in}}\\
   Saha Institute of Nuclear Physics,\\ 1/AF Bidhan  
        Nagar, Kolkata 700064, India }
\date{}

\begin{document}

\maketitle

\begin{abstract}
  Continuous $U(1)_{L_\mu-L_\tau}$ symmetry can generate quasi
  degenerate mass spectrum for both left handed light and right handed
  heavy Majorana neutrinos assuming that the symmetry preserving non
  zero parameters are nearly same. There is an accidental $\mu\tau$
  exchange symmetry in the light and heavy neutrino Majorana mass
  terms. This implies $\theta_{13}=0$ and $\theta_{23}=\frac{\pi}{4}$.
  In addition it generates another zero mixing angle and one zero mass
  difference. We restrict ourselves to type-I See-Saw mechanism for
  generation of light neutrino mass. We have found that under
  $U(1)_{L_\mu-L_\tau}$ symmetry cosmological lepton asymmetry
  vanishes. We break $U(1)_{L_\mu-L_\tau}$ such a way that the
  $\mu\tau$ exchange symmetry preserves in the neutrino sector. We
  have seen that light neutrino phenomenology can be explained under
  soft breaking of this symmetry. We have observed that softness of
  this symmetry breaking depends on the degeneracy of the light
  neutrino mass spectrum.  Quasi-degeneracy of right handed neutrino
  mass spectrum opens an option for resonant leptogenesis.  The
  degeneracy of the right handed neutrino mass spectrum is restricted
  through light neutrino data. We observed that for generation of
  right sized baryon asymmetry common neutrino mass scale $m_0$ have
  to be of the order of $\sqrt{\Delta m^2_{\rm atm}}$ and
  corresponding right handed neutrino mass scale have to be nearly
  $10^{13}$ GeV. We also have discussed the effect of RG evolution on
  light neutrino spectrum and also on baryon asymmetry.

\end{abstract}
PACS number(s): 14.60.Pq, 11.30.Hv, 98.80.Cq

\section{Introduction}
Neutrinos are massless particles in the Standard Model(SM). So, there
is no option for leptonic mixing. Astronomical observations of
neutrino oscillation in the last ten years have confirmed that
neutrinos have masses and they can mix. Neutrino oscillation data
provide us the information about the mass-square differences of the
neutrinos and the leptonic mixing angles.

Solar neutrino \cite{Aharmim:2005gt,solar} and reactor neutrino
\cite{kl2} oscillation data analysis \cite{analyzes, Strumia:2005tc}
provide us large but non maximal solar mixing angle
$\theta_\odot\equiv \theta_{12} \sim 34^\circ$ and small solar mass square
difference $\Delta m_\odot^2\equiv \Delta m^2_{21}=m_2^2-m_1^2\sim
8\times 10^{-5}{\rm eV}^2 $.
Atmospheric neutrino \cite{SKatm} and accelerator neutrino oscillation
data \cite{k2k} analysis give almost maximal atmospheric mixing
angle $\theta_{\rm atm}\equiv\theta_{23} \sim 45^\circ$ and larger atmospheric
mass square difference $\Delta m_{\rm atm}^2\equiv|\Delta
m^2_{32}|=|m_3^2- m^2_2|\sim 2.1\times 10^{-3}{\rm eV}^2$ compare to solar mass square difference.  The
CHOOZ experiment \cite{Apollonio:2002gd} gives only a upper limit on
$\theta_{13}$ \cite{Strumia:2005tc}, $\theta_{13}<10^\circ$. Here the
mixing angles are in the standard parametrization of PMNS matrix
\cite{PMNS}.

Apart from neutrino oscillation data there are neutrinoless double
beta($0\nu\beta\beta$) decay experimental data \cite{Elliott:2004hr}
and WMAP data \cite{Spergel:2003cb} about neutrino mass directly.
Majorana character (particle is its own antiparticle) of neutrino gives an option for $0\nu\beta\beta$ decay.
It provides direct bound on light neutrino mass scale
\cite{Fogli:2004as}
\begin{eqnarray}
m_{eff} \equiv
\left|(m_\nu)_{ee}\right|=\left|\sum_{i=1}^3 U_{ei}^2 m_i\right| < 0.89
{\rm eV}\quad{\rm at ~1\sigma}.
\label{nldb}
\end{eqnarray}
On the other hand WMAP data gives bound on the sum of three generation
masses of neutrinos. This bound is
\begin{eqnarray}
\sum_{i=1}^3m_i\le 0.69 {\rm eV}\quad{\rm at ~95\%~ confidence ~level}.
\label{wmap1}
\end{eqnarray}
It is necessary to point out that atmospheric neutrino data can not
settle the sign of $\Delta m^2_{32}$. This sign ambiguity generates
two kinds of hierarchical mass spectrum. Those are normal hierarchical
$m_1\ll m_2\ll m_3$ and inverted hierarchical $m_1< m_2\gg m_3$. For
$m_{i,j}\gg \sqrt{|\Delta m^2_{ij}|}$ we can have quasidegenerate mass
spectrum, $m_1\approx m_2\approx m_3$. In this case both normal and
inverted ordering are also possible due to the same sign ambiguity of
$\Delta m^2_{32}$.  So, from neutrino oscillation data we have
\begin{eqnarray}
&&m_2=m_0\qquad m_1=\sqrt{m_0^2-\Delta m_\odot^2}\qquad m_3=\sqrt{m_0^2\pm\Delta
m_{\rm atm}^2}.
\label{neutos2}
\end{eqnarray}
Now if we use these to WMAP bound we have  
\begin{eqnarray}
&&m_0+\sqrt{m_0^2-\Delta m_\odot^2}+\sqrt{m_0^2\pm\Delta
m_{\rm atm}^2}\le 0.69\rm{eV}\nonumber\\
&&\qquad\Longrightarrow m_0\le 0.23\rm{eV}.
\label{wmap2}
\end{eqnarray}
If $m_0$ remains in the range $\sqrt{\Delta m_{\rm atm}^2}\le m_0\le
0.23\rm{eV}$ then hierarchical nature of the neutrino mass will no
longer exist and we can have the degenerate kind of mass spectrum with
both type of ordering. In this article we want to study a flavor
symmetry which leads to this kind of light neutrino mass spectrum
along with the two large mixing angle considering one mixing angle
zero. 

Now we will briefly point out the existing flavor symmetric models.
$\mu\tau$ exchange symmetry (permutation symmetry
$S_2^{\mu\tau}$)\cite{mutau,Grimus:2003sf,Mohapatra:1998rq,\cite{Aizawa:2004qf}} in the
neutrino sector generates following Majorana-like neutrino mass
matrix,
\begin{eqnarray}
m_\nu=\left(\begin{array}{ccc} A & B & B\\ B & C & D\\ B & D & C
  \end{array}\right).
\label{perm}
\end{eqnarray}
It explains one zero mixing angle ($\theta_{13}$) and one maximal
mixing angle ($\theta_{23}$). Other angle and all masses of light
neutrinos are unconstrained. To gain more predictability various kinds
of $U(1)$ flavor symmetry can be imposed. Among them three are
phenomenologically important. Those are $U(1)_{L_e}$
\cite{Choubey:2004hn,\cite{Aizawa:2004qf}}, $U(1)_{L_e-L_\mu-L_\tau}$
\cite{lelmlt, grimus, Petcov:2004rk} and $U(1)_{L_\mu-L_\tau}$
\cite{Choubey:2004hn}. We will call them as ${L_e}$,
$L_e-L_\mu-L_\tau$ and $L_\mu-L_\tau$ symmetry. These three flavor
symmetries can explain three different kinds of neutrino mass
spectrum. Normal and inverted hierarchical mass spectrum can be
obtained under ${L_e}$ and $L_e-L_\mu-L_\tau$ symmetry respectively.
But, $L_\mu-L_\tau$ symmetry opens an option for degenerate kind of
mass spectrum. It generates the following Majorana-like neutrino mass
matrix
\begin{eqnarray}
\qquad m_\nu=\left(\begin{array}{ccc} A & 0 & 0\\ 0 & 0 & D\\ 0 & D & 0
  \end{array}\right)
\end{eqnarray}
which gives $m_2=m_3$, $\theta_{13}=\theta_{12}=0$ and
$\theta_{23}=\pi/4$.  This is the special form of Eq.\ (\ref{perm})
with $B=C=0$. If we assume $|A|\approx |D|$, we have $m_2=m_3\approx m_1$
which corresponds to ``quasi degenerate'' mass spectrum.

Discrete flavor symmetry $A_4$ \cite{A4} has an ability to produce
degenerate kind of light neutrino mass spectrum. We concentrate only
to the continuous $U(1)$ flavor symmetry $L_\mu-L_\tau$ in type-I see
saw model with three generations of right handed neutrino. In this
respect, it has also been shown that the flavor symmetry $L_\mu -
L_\tau$ is (in the SM) anomaly free and may be gauged \cite{gauge}. In
\cite{Choubey:2004hn} the authors make some general study of this
$L_\mu-L_\tau$ symmetry breaking under democratic as well as
anarchical perturbation over $L_\mu-L_\tau$ symmetric neutrino mass
matrix along with the radiative corrections.  They also pointed out
that such kind of mass matrices can be generated through type-I
see-saw mechanism. We disregard the radiative correction in our
analysis. We will comment about it at the end. The $L_\mu-L_\tau$
symmetric Majorana mass terms of neutrino (both left and right handed)
shows permutation symmetry $S_2^{\mu\tau}$ .  To keep $\theta_{13}=0$
and $\theta_{23}=\pi/4$, we break $L_\mu-L_\tau$ symmetry keeping
$S_2^{\mu\tau}$ symmetry in the neutrino sector only.  $L_\mu-L_\tau$
symmetry generate quasi-degenerate mass spectrum both in right and
left handed neutrino under the assumption that the symmetry preserving
non-zero parameters are nearly equal. It opens an option for resonant
kind of leptogenesis. Present WMAP \cite{Spergel:2003cb} prediction of
baryon asymmetry (baryon density to photon density ratio) is
\begin{eqnarray}
\eta_B=(6.1^{+0.3}_{-0.2})\times 10^{-10}.
\label{bad}
\end{eqnarray}
We will see that $L_\mu-L_\tau$ symmetry forbids leptogenesis. We have
tried to study whether it is possible or not to generate light
neutrino phenomenological results and the right sized baryon asymmetry
through leptogenesis under soft breaking of $L_\mu-L_\tau$ symmetry.

In the Section.  \ref{symmetry} we will introduce $L_\mu-L_\tau$
symmetry to the Lagrangian relevant for our discussion and will show
the structures of Dirac-type mass matrix and right handed neutrino
mass matrix. In the subsequent two subsections of this section will
consist of the study of light neutrino phenomenology and leptogenesis
when $L_\mu-L_\tau$ symmetry is preserved. In the next Section.
\ref{ltb} we will break the $L_\mu-L_\tau$ symmetry. In its first
Subsection \ref{lnp} we will discuss about the possibility of
generation of experimental values of light neutrino masses and mixing
angles after soft breaking of $L_\mu-L_\tau$ symmetry. In the next
Subsection.\ref{abl} we will study whether the broken $L_\mu-L_\tau$
symmetry can generate right sized lepton asymmetry for predicting WMAP
data of baryon asymmetry or not. We will discuss about the RG effects on
our results at the end of this subsection. Then we will conclude in
the last Section. \ref{conclu}.
 
\section{$\tilde L={L_\mu}-{L_\tau}$ symmetry}
\label{symmetry}
For convenience from now we will call $L_\mu-L_\tau$ symmetry as
$\tilde L$ symmetry. Before going to the details of this symmetry we
want to introduce the Lagrangian relevant for our discussion. We have
introduced three generations of $SU(2)_L\times U(1)_Y$ singlet right
handed neutrinos in addition to SM particles.
The relevant part  of $SU(2)_L\times U(1)_Y$ symmetric Lagrangian  for
our discussion is
\begin{eqnarray}
-\mathscr L^{Y}_{l}={{\bar{\psi_l}}}_L(Y^l)_{ll'}
\Phi{l}^\prime_R+
{{\bar{\psi_l}}}_L(Y^{\nu})_{ll'}\widetilde\Phi{N_{l'}}_R
+\frac{1}{2}{\bar{ N_l}^c}_L(M_R)_{ll'}{N_{l'}}_R +h.c.
\label{lagy}
\end{eqnarray}
where
\begin{eqnarray}
{\psi}_{lL}=\left(\begin{array}{c}{\nu_l}_L\\l_L\end{array}\right)\quad\Phi=\left(\begin{array}{c}\phi^+\\\phi^0\end{array}\right),
\label{doblets}
\end{eqnarray}
$\widetilde\Phi = i\sigma_2\Phi^*$, $N^c_L=(N_R)^c=-\gamma_0CN_R^*$
and $l,l'=e,\mu,\tau$. $\nu_{lL}$ and $N_{lR}$ are the left and right
handed neutrinos. $\Phi$ is standard scalar doublet containing charged
scalar $\phi^+$ and neutral scalar$\phi^0$.  After $SU(2)_L\times
U(1)_Y$ symmetry breaking to $ U(1)_Q$ mass terms are generated
through vacuum expectation values of
$\Phi$ : $\left<\Phi\right>=\left(\begin{array}{c} 0\\
    \frac{v}{\sqrt{2}}\end{array}\right)$. The mass terms are
\begin{eqnarray}
-\mathscr{L}^m={\bar l}_L(m_l)_{ll'} {l}^\prime_R+
{{\bar{\nu_l}}}_L(m_D)_{ll'}{N_{l'}}_R
+\frac{1}{2}{\bar{ N_l}^c}_L(M_R)_{ll'}{N_{l'}}_R +h.c.
\label{lagml}
\end{eqnarray}
where $m_l=Y^lv/\sqrt{2}$
charged lepton mass matrix,
$m_D=Y^\nu v/\sqrt{2}$ Dirac type neutrino mass matrix and $M_R$
 is the bare Majorana type right handed neutrino mass matrix. Without loss
 of generality we consider a basis where $m_l$ is diagonal, $m_l={\rm
   diag}(m_e, m_\mu, m_\tau)$. Remaining part of the mass Lagrangian
 Eq.\ (\ref{lagml}) gives neutrino mass terms in the following form
\begin{eqnarray}
-\mathscr{L}^m_\nu=\frac{1}{2}\left(\begin{array}{cc}\bar{\nu}_L &
    {\bar N}^c_L\end{array}\right)_l\left(\begin{array}{cc} 0 &
      m_D \\ m_D^T & M_R\end{array}\right)_{ll'}\left(\begin{array}{c}\nu^c_R\\
     N_R\end{array}\right)_{l'}+h.c.
\label{lagn}
\end{eqnarray}
using the general identity 
\begin{eqnarray}
{\bar \psi_L}m\psi'_R=\bar{\psi^\prime}^c_Lm^T\psi^c_R.
\label{ident}
\end{eqnarray}
For $O(M_R)\gg O(m_D)$ we obtained effective Majorana-like light neutrino mass term
\begin{eqnarray}
-\mathscr{L}^m_{\nu_L}=\frac{1}{2}\bar{\nu_l}_L(m_\nu)_{ll'}{\nu_l}^c_R
\label{lag4}
\end{eqnarray}
where
\begin{eqnarray}
m_\nu=-m_DM_R^{-1}m_D^T.
\label{see-saw}
\end{eqnarray}
Eq.\ (\ref{see-saw}) is known as See-Saw formula which explains
lightness of neutrino.

Now let us see the  $\tilde{L}$ number of the different leptonic fields:
\begin{eqnarray}
&& e,\nu_e, N_e\rightarrow\qquad \tilde{L}=0\nonumber\\
&&\mu,\nu_\mu, N_\mu\rightarrow\qquad \tilde{L}=+1\nonumber\\
&&\tau,\nu_\tau, N_\tau\rightarrow\qquad \tilde{L}=-1.
\label{ltcharge}
\end{eqnarray}
All other fields have zero $\tilde{L}$ number.  Now we want to impose
$\tilde{L}$ symmetry to the Lagrangian. $\tilde{L}$ symmetry
simultaneously diagonalizes $m_l$ and $m_D$. Under this symmetry $m_D$
and $M_R$ take the following structure \cite{Choubey:2004hn}
\begin{eqnarray}
m_D=\left(\begin{array}{ccc} a & 0 & 0\\ 0 & b & 0\\ 0 & 0 & c
  \end{array}\right) \qquad
M_R=\left(\begin{array}{ccc} p & 0 & 0\\ 0 & 0 & q\\ 0 & q & 0
  \end{array}\right).
\label{matform}
\end{eqnarray}
Without loss of generality we choose that the elements of $m_D$ are
real positive. So $a$, $b$ and $c$ are real positive. $p$ and $q$ are
complex in general.
\subsection{Neutrino mass and mixing}
\label{nmasmix}
Using the
 $\tilde L$ symmetric form of $m_D$ and $M_R$ from Eq.\
 (\ref{matform}) in the see saw formula in Eq.\
(\ref{see-saw}) we have 
\begin{eqnarray}
m_\nu=-\left(\begin{array}{ccc} \frac{a^2}{p} & 0 & 0\\ 0 & 0
    &\frac{bc}{q}\\ 0 &\frac{bc}{q} & 0
  \end{array}\right).
\label{lnmas}
\end{eqnarray}
Both $M_R$ and $m_\nu$ shows the permutation symmetry $S_2^{\mu\tau}$
which is an accidental symmetry here.  $m_\nu$ can be diagonalized as
\begin{eqnarray}
U^\dagger m_\nu U^*={\rm diag}(m_1, m_2, m_3)
\end{eqnarray}
where
\begin{eqnarray}
m_1=\left|\frac{a^2}{p}\right|\qquad m_2=m_3=\left|\frac{bc}{q}\right|
\label{nutmas}
\end{eqnarray}
are the light neutrino masses. $U$ relates  the flavor
basis to  eigen basis of light neutrino:
\begin{eqnarray}
\nu_{lL}=\sum_{i=1}^{3}U_{li}\nu_{iL}.
\end{eqnarray}
This $U$ is the leptonic mixing matrix. The form of $U$ we obtain 
\begin{eqnarray}
U=\left(\begin{array}{ccc} ie^{-i\theta_p/2} & 0 & 0\\ 0 & ie^{-i\theta_q/2}/\sqrt{2} & e^{-i\theta_q/2}/\sqrt{2} \\ 0 & ie^{-i\theta_q/2}/\sqrt{2} & -e^{-i\theta_q/2}/\sqrt{2} 
\end{array}\right)
\label{ufm}
\end{eqnarray}
where $\theta_p=\arg(p)$ and $\theta_q=\arg(q)$. $U$ of Eq.\ 
(\ref{ufm}) can be rewritten as
\begin{eqnarray}
U=U_{p_1}OU_{p_2}=\left(\begin{array}{ccc}\exp^{-i\theta_p/2} & 0 & 0\\ 0 & \exp^{-i\theta_q/2} & 0 \\ 0 & 0 & -\exp^{-i\theta_q/2} 
\end{array}\right)\left(\begin{array}{ccc} 1 & 0 & 0\\ 0 & \frac{1}{\sqrt{2}}
  & \frac{1}{\sqrt{2}} \\ 0 & \frac{-1}{\sqrt{2}} & \frac{1}{\sqrt{2}} 
\end{array}\right)\left(\begin{array}{ccc}  i & 0 & 0\\ 0 &  i & 0 \\
  0 & 0 & 1
\end{array}\right).
\label{ufm1}
\end{eqnarray}
 The phases $U_{p_1}$
can be absorbed to the charged lepton fields in charged current
interaction. Non absorbable Majorana phases are in $U_{p_2}$. Matrix
$O$ gives the mixing angles. So we have the mass-squared differences
\begin{eqnarray}
\Delta m_{32}^2=0\qquad \Delta m_{21}^2=\left|\frac{bc}{q}\right|^2-\left|\frac{a^2}{p}\right|^2
\label{massd}
\end{eqnarray}
and the mixing angles
\begin{eqnarray}
\theta_{12}=\theta_{13}=0\qquad \theta_{23}=\frac{\pi}{4}
\label{angl}
\end{eqnarray}
For $a\approx b\approx c$ and $|p|\approx |q|$ we can have nearly
degenerate mass spectrum. Recall that $\theta_{12}=0$ is not a
consequence of $S_2^{\mu\tau}$ symmetry: it is the additional constraint
coming from the $\tilde{L}$ symmetry.
\subsection{Leptogenesis}
\label{lpt}

Let us briefly discuss about right handed Majorana neutrino decay
generated leptogenesis. There is a Dirac type Yukawa interaction in
Eq.\ (\ref{lagy}) of right handed neutrino (${N}_l$) with SM lepton
doublet and Higgs doublet. At the energy scale where $SU(2)_L\times
U(1)_Y$ symmetry is preserved, physical right handed neutrino ${N}_i$
with definite mass can decay both to charged lepton with charged
scalar and light neutrino with neutral scalar. Due to the Majorana
character of ${N}_i$ conjugate process is also possible. If out of
equilibrium decay of ${N}_i$ in conjugate process occur at different
rate from actual process, net lepton number will be generated. The CP
asymmetry of decay is characterized by a parameter $\varepsilon_i$
which is defined as
\begin{eqnarray}
\varepsilon_i&=&\frac{\Gamma_{{N}_i\rightarrow
    l^-\phi^+,\nu_l\phi^0}-\Gamma_{{N}_i\rightarrow
    l^+\phi^-,\nu_l^c\phi^{0*}}}{\Gamma_{{N}_i\rightarrow
    l^-\phi^+,\nu_l\phi^0}+\Gamma_{{N}_i\rightarrow
    l^+\phi^-,\nu_l^c\phi^{0*}}}.
\label{cpasym}
\end{eqnarray}
At $SU(2)_L\times U(1)_Y$ symmetry preserving scale we can have
exactly the masses of the right handed neutrinos diagonalising only
the bare mass matrix $M_R$ in Eq.\ (\ref{matform}). We diagonalize
$M_R$ as
\begin{eqnarray}
V^\dagger M_R V^*={\rm diag}(M_1, M_2, M_3)
\label{mrd}
\end{eqnarray}
where this $V$ relates the flavor basis to eigen basis of right handed
neutrino:
\begin{eqnarray}
N_{lR}=\sum_{i=1}^{3}V_{li}^*N_{iR}.
\label{fmr}
\end{eqnarray}
and the $M_i$ are the masses of right handed neutrinos.  Now starting
with the general resummed effective Yukawa couplings in one loop for
three generation of right handed neutrino from the Appendix-A Eq.  A.7
in the Ref:\cite{Pilaftsis:2003gt} we obtain the two parts of
$\varepsilon_i$, the general vertex part
\begin{eqnarray}
\varepsilon_i^V=\frac{1}{4\pi v^2h_{ii}}\sum_{j\ne
  i}Im(h_{ij}^2)\times \left[\sqrt{x_{ij}}\left\{1-(1+x_{ij})\ln(1+\frac{1}{x_{ij}})\right\}\right]
\label{vertex}
\end{eqnarray}
and the self energy part 
\begin{eqnarray}
\varepsilon_i^S=\frac{1}{4\pi v^2h_{ii}}\sum_{j\ne
  i}Im(h_{ij}^2)\times\left[\frac{(1-x_{ij})\sqrt{x_{ij}}}{(1-x_{ij})^2+(h_{jj}/4\pi v^2)^2}\right].
\label{self}
\end{eqnarray}
where $x_{ij}=M_j^2/M_i^2$ and
\begin{eqnarray}
h=m_D'^\dagger m_D',
\label{h}
\end{eqnarray}
where 
\begin{eqnarray}
m_D'=m_DV^*.
\label{mdp}
\end{eqnarray}
The vertex part $\varepsilon_i^V$ in Eq.\ (\ref{vertex}) has the
standard form \cite{Fukugita:1986hr,leptogen}.  The self energy
$\varepsilon_i^S$ part in Eq.\ (\ref{self}) is true for general two
generation right handed neutrino \cite{Pilaftsis:2003gt}. Complicated
expression for three generation self energy part $\varepsilon_i^S$ can
be reduced to the expression in Eq.\ (\ref{self}) for small Yukawa
couplings or under some special conditions. One condition is that any
one of the right handed neutrino has very larger value of mass compare to
other two \cite{Pilaftsis:2003gt}. We observed that for any one
generation $k$ if $h_{kj}=0$ for all $j\ne k$, then the expression in
Eq.\ (\ref{self}) also can be used for three generation as well.  It
turns out to the fact that corresponding $\varepsilon_k=0$. Other two
generations can give nonzero $\varepsilon_i$ ($i\ne k$), if
$Im(h_{ij}^2)\ne 0$ and function in the parenthesis of Eq.\ 
(\ref{vertex}) and Eq.\ (\ref{self}) are nonzero. This whole
discussion about the expression of $\varepsilon_i$ is done keeping in
mind that the right handed neutrinos can have masses with very close
in value. There is a possibility in that case where the two quantities
of the denominator in the self energy expression $\varepsilon_i^S$ of
Eq.\ (\ref{self}) become comparable in value. This is equivalent to
the resonant condition $M_i-M_j \approx\Gamma_{i,j}/2$ where
$\Gamma_i$ is the decay width of $N_i$. Then the the self energy
contribution $\varepsilon_i^S$ will be enhanced compare to
$\varepsilon_i^V$. This is so called resonant leptogenesis. For
hierarchical mass spectrum of right handed neutrino the second term in
the denominator of $\varepsilon_i^S$ Eq.\ (\ref{self}) for the self
energy contribution become negligible and it can be reduced to the
usual self energy function $\sqrt{x_{ij}}/(1-x_{ij})$ \cite{leptogen}.

Now coming back to our original discussion for this model we first
state the results of diagonalization of $\tilde{L}$ symmetric $M_R$ of
Eq.\ (\ref{matform}) as in Eq.\ (\ref{mrd}). The obtained masses of
the right handed neutrinos are
\begin{eqnarray}
M_1=\left|p\right|\qquad M_2=M_3=\left|q\right|.
\label{rnutmas}
\end{eqnarray}
The form of diagonalizing matrix $V$ of Eq.\ (\ref{mrd}) is
\begin{eqnarray}
V=\left(\begin{array}{ccc} e^{i\theta_p/2} & 0 & 0\\ 0 & e^{i\theta_q/2}/\sqrt{2} & -ie^{i\theta_q/2}/\sqrt{2} \\ 0 & e^{i\theta_q/2}/\sqrt{2} & ie^{i\theta_q/2}/\sqrt{2} 
\end{array}\right).
\label{vform}
\end{eqnarray}
Using this $V$ we can easily get $m_D'$ from Eq.\ (\ref{mdp}). Putting
this $m_D'$ to Eq.\ (\ref{h}) we can have the relevant matrix $h$ for the
calculation of $\varepsilon_i$. So, the obtained form of $h$ is
\begin{eqnarray}
h&=&m_D'^\dagger m_D'\nonumber\\
&=&\left(\begin{array}{ccc} a^2 & 0 & 0\\ 0 &
    (b^2+c^2)/2 &
    -i(b^2-c^2)/2 \\ 0 &
    i(b^2-c^2)/2 & (b^2+c^2)/2 
\end{array}\right).
\label{hmat}
\end{eqnarray}
$h_{12}=h_{13}=0$ ensure that $\varepsilon_1=0$. So, $N_1$ decay can
not generate lepton asymmetry. Remaining two generation are degenerate
in mass. As $x_{23}=1$ self energy part $\varepsilon_{2,3}^S$ as in
Eq.\ (\ref{self}) does not contribute to $\varepsilon_{2,3}$. But,
as $Im(h_{23}^2)=0$ both $\varepsilon_{2}$ and $\varepsilon_{3}$
vanishes individually.  This is the artifact of the $\tilde{L}$
symmetry. So, lepton asymmetry cannot be generated when $\tilde{L}$
symmetry is exact. Without calculating decay asymmetry parameters one
can check CP is preserved or not.  We want to calculate a weak
basis invariant CP odd quantity $\Delta_{\rm CP}$ which have to be
nonzero in order to generate non zero lepton asymmetry. $\Delta_{\rm CP}$
is defined as \cite{Branco:2001pq}
\begin{eqnarray}
\Delta_{\rm CP}\equiv{\rm ImTr}\left[m_D^\dagger m_DM_R^\dagger
  M_RM_R^\dagger m_D^Tm_D^*M_R\right].
\end{eqnarray}
In the physical mass basis of right handed neutrino it takes the form \cite{Branco:2001pq}
\begin{eqnarray}
\Delta_{\rm CP}&=&M_1M_2(M_2^2-M_1^2){\rm Im}(h_{12}^2)+M_1M_3(M_3^2-M_1^2){\rm Im}(h_{13}^2\nonumber)\\
&&+M_2M_3(M_3^2-M_2^2){\rm Im}(h_{23}^2).
\label{dcp}
\end{eqnarray}
$h_{12}=h_{13}=0$ and ${\rm Im}(h_{23}^2)=0$ imply that
$\Delta_{\rm CP}=0$. So, it explains the reason for vanishing of lepton
asymmetry under $\tilde L$ symmetry. Before closing our discussion for
this subsection we want to point out about the imposition of
$S_2^{\mu\tau}$ symmetry in the neutrino sector only. Under
$S_2^{\mu\tau}$ symmetry we have $b=c$ of the elements of $m_D$ in
Eq.\ (\ref{matform}). It makes $h$ of Eq.\ (\ref{hmat}) exactly
diagonal. We will see that whether it is possible or not to generate
lepton asymmetry along with light neutrino phenomenology under soft
violation of $\tilde{L}$ symmetry keeping $S_2^{\mu\tau}$ unaltered.
\section{Soft breaking of $\tilde{L}$ symmetry}
\label{ltb}
As $\tilde{L}$ symmetric $M_R$ and $m_\nu$ both shows the
$S_2^{\mu\tau}$ symmetry, we break $\tilde{L}$ symmetry
such a way that $S_2^{\mu\tau}$ symmetry preserves. We break the
$\tilde{L}$ symmetry only in the right handed neutrino bare mass
matrix $M_R$, preserving $S_2^{\mu\tau}$ and keeping the $\tilde{L}$
symmetry everywhere else for minimality. To see this $S_2^{\mu\tau}$
symmetry also in $m_\nu$ we have to impose this symmetry to $m_D$ also
which implies $b=c$ in $m_D$ matrix of Eq.\ (\ref{matform}). Now let
us write down the $M_R$ and $m_D$ including $\tilde{L}$ violating
effects
\begin{eqnarray}
m_D=\left(\begin{array}{ccc} a & 0 & 0\\ 0 & b & 0\\ 0 & 0 & b
  \end{array}\right) \qquad
M_R=\left(\begin{array}{ccc} p & \epsilon & \epsilon \\ \epsilon & \epsilon' & q\\ \epsilon & q & \epsilon'
  \end{array}\right).
\label{sbmdmr}
\end{eqnarray}
Here, $\epsilon$ and $\epsilon'$ are the $\tilde{L}$ symmetry violating
small parameters and they are complex in general. Seeing also in the
$\tilde{L}$ symmetric $\Delta m_{21}^2$ in Eq.\ (\ref{massd}) we
require some assumptions which are $|p|\approx |q|$ and $b\approx
a$. We quantify these assumptions as
\begin{eqnarray}
|q|=|p|(1+\Delta')\qquad {\rm and} \qquad b=a(1+\Delta)
\label{assum}
\end{eqnarray}
for using in future. Here $\Delta$ and $\Delta'$ are the small
numbers. This will generate quasi-degenerate mass spectrum for both in
light and heavy neutrino sector. We also want to define three
quantities which we will use in later time. Those are
\begin{eqnarray}
\delta=\frac{\epsilon}{|p|}{\rm,}\qquad
\delta'=\frac{\epsilon'}{|p|}\quad {\rm and}\quad m_0=\frac{a^2}{|p|},
\label{ddpm0}
\end{eqnarray}
where $\delta$, $\delta'$ are the measures of $\tilde{L}$ violation
and $m_0$ is the common light neutrino mass scale. Now we want to
study whether it is possible or not to generate right sized baryon
asymmetry through lepton asymmetry along with the successful low energy
light neutrino phenomenology.
\subsection{Light neutrino phenomenology}
\label{lnp}
Using  $\tilde{L}$ symmetry violating $m_D$ and $M_R$ in see-saw formula
in Eq.\ (\ref{see-saw}) we have the following form of light neutrino
mass matrix
\begin{eqnarray}
m_\nu=\left(\begin{array}{ccc} A & B & B\\ B & C & D\\ B & D & C
  \end{array}\right).
\label{mnu}
\end{eqnarray}
which is in general $S_2^{\mu\tau}$ symmetric form. The elements are
\begin{eqnarray}
A=-\frac{(q+\epsilon')a^2}{F}\qquad B=\frac{ab\epsilon}{F} \qquad
C=-\frac{b^2(\epsilon^2-p\epsilon')}{(q-\epsilon')F} \qquad D=\frac{b^2(\epsilon^2-pq)}{(q-\epsilon')F}
\label{mnuelem}
\end{eqnarray}
where 
\begin{eqnarray}
F=(pq+p\epsilon'-2\epsilon^2).
\label{sclm0}
\end{eqnarray}
$B$ and $C$ should be small because the $\tilde{L}$ symmetry is
approximate. The dependence of $B$ and $C$ on the symmetry violating
parameters $\epsilon$ and $\epsilon'$ ensure it. Next discussions are
similar to $S_2^{\mu\tau}$ symmetric
\cite{mutau,Grimus:2003sf,Mohapatra:1998rq} case. Smallness of $B$ and
$C$ is the special feature here. In the process of finding masses and
mixing we first diagonalize $H$ which is defined
\begin{eqnarray}
H=m_\nu m_\nu^\dagger=\left(\begin{array}{ccc} P & Q^* &  Q^* \\ Q & R & S
    \\ Q & S & R\end{array}\right)
\end{eqnarray}
where
\begin{eqnarray}
P = \left|A\right|^2 + 2\left|B\right|^2 &\qquad& Q
=A^*B+B^*(C+D)\nonumber\\
R=\left|B\right|^2 + \left|C\right|^2 + \left|D\right|^2 &\qquad&
S=\left|B\right|^2 + DC^* + CD^*
\label{PQRS}
\end{eqnarray}
We diagonalize $H$ as
\begin{eqnarray}
U^{\dagger}HU={\rm diag}(m_1^2, m_2^2, m_3^2)
\label{hdiag}
\end{eqnarray}
where the eigen values are \cite{mutau,Grimus:2003sf,Mohapatra:1998rq}
\begin{eqnarray}
m_{2,1}^2=\frac{P+R+S\pm\sqrt{(R+S-P)^2+8\left|Q\right|^2}}{2}\qquad
m_3^2 = R - S
\label{eigen}
\end{eqnarray}
and the mixing matrix
\begin{eqnarray}
U=\left(\begin{array}{ccc} J & \frac{\sqrt{2}Q^*J}{m_2^2-P} & 0 \\
-\frac{QJ}{m_2^2-P} & \frac{J}{\sqrt{2}} & -\frac{1}{\sqrt{2}} \\
-\frac{QJ}{m_2^2-P} & \frac{J}{\sqrt{2}} &
\frac{1}{\sqrt{2}}\end{array}\right)\quad {\rm where}\quad J=\sqrt{\frac{m_2^2-P}{m_2^2-m_1^2}}.
\label{mixmat1}
\end{eqnarray}
This matrix $U$ can diagonalize $m_\nu$ but the entries may not be
real positive. So, additional diagonal phase matrix have to multiply
in the right side of the above $U$. We are not interested about this
phase matrix in the analysis of light neutrino phenomenology. We will see
that such phase matrix play an important role in lepton asymmetry
generation. We can rewrite $U$ in Eq.\ 
(\ref{mixmat1}), and we have
\begin{eqnarray}
U=\left(\begin{array}{ccc}\exp^{i\theta_Q}  & 0 & 0 \\
0 & 1 & 0 \\
0 & 0 & -1 \end{array}\right)\left(\begin{array}{ccc} J & \frac{\sqrt{2}|Q|J}{m_2^2-P} & 0 \\
-\frac{|Q|J}{m_2^2-P} & \frac{J}{\sqrt{2}} & \frac{1}{\sqrt{2}} \\
\frac{|Q|J}{m_2^2-P} & -\frac{J}{\sqrt{2}} & \frac{1}{\sqrt{2}}\end{array}\right)\left(\begin{array}{ccc}\exp^{-i\theta_Q}  & 0 & 0 \\
0 & 1 & 0 \\
0 & 0 & -1 \end{array}\right)
\label{up}
\end{eqnarray}
where $\theta_Q=\arg(Q)$. The matrix in the middle of the above
decomposition of $U$ gives the mixing angle and it exactly matches
with the $O$ matrix in Eq.\ (\ref{ufm1}) in exact $\tilde{L}$
symmetric case, $Q=0$ and $J=1$. So, we have the mixing
pattern as follow \cite{mutau,Grimus:2003sf,Mohapatra:1998rq}
\begin{eqnarray}
\theta_{13}&=&0 \qquad \theta_{23}=\pi/4\nonumber\\
\tan{2\theta_{12}}&=& \frac{2\sqrt{2}|Q|}{R+S-P}.
\label{mixpat}
\end{eqnarray}
Now from the Eq.\ (\ref{eigen}) we have the mass differences
\begin{eqnarray}
\Delta m^2_{21}=\sqrt{(R+S-P)^2+8\left|Q\right|^2}=(R+S-P)\sqrt{1+\tan^22\theta_{12}}=2\sqrt{2}|Q|\sqrt{\frac{1+\tan^22\theta_{12}}{\tan^22\theta_{12}}}
\label{m12}
\end{eqnarray}
and
\begin{eqnarray}
\Delta m^2_{32}=-\frac{\Delta
  m^2_{21}}{2}\left[1-\frac{1}{\sqrt{1+\tan^22\theta_{12}}}\right]-2S. 
\label{m23}
\end{eqnarray}
These are the results from general $S_2^{\mu\tau}$ symmetric case. Now
substituting the expression of  $A$, $B$, $C$ and $D$ from Eq.\ 
(\ref{mnuelem}) to $P$, $Q$, $R$ and $S$ of Eq.\
(\ref{PQRS})  we get the following approximate results keeping terms
first order in $\delta$, $\delta'$, $\Delta$ and $\Delta'$
\begin{eqnarray}
&&R+S-P\approx 2m_0^2(2\Delta-\Delta'-|\delta'|\cos\gamma)
\qquad Q\approx-2m_0^2|\delta|\cos\frac{\rho}{2}\exp^{i\theta_Q}\nonumber\\
&&S\approx-2m_0^2\delta'\cos\gamma.
\label{params}
\end{eqnarray}
which give us
\begin{eqnarray}
\tan 2\theta_{12}&\approx &\frac{2\sqrt{2}|\delta|\cos\frac{\rho}{2}}{2\Delta-\Delta'-|\delta'|\cos\gamma}\nonumber\\
\Delta m_{21}^2&\approx &
2m_0^2(2\Delta-\Delta'-|\delta'|\cos\gamma)\sqrt{1+\tan^22\theta_{12}}\nonumber\\
&\approx &
4\sqrt{2}m_0^2|\delta|\cos\frac{\rho}{2}\sqrt{\frac{1+\tan^22\theta_{12}}{\tan^22\theta_{12}}}\nonumber\\
\Delta m_{32}^2&\approx &-\frac{\Delta m_{21}^2}{2}\left[1-\frac{1}{\sqrt{1+\tan^22\theta_{12}}}\right] +4m_0^2|\delta'|\cos\gamma.
\label{predic}
\end{eqnarray}
where the phase of the approximated $Q$ is $2\theta_Q=\arg{(pq^*)}$, $\rho=\arg{(\epsilon^{2}q^*p^*)}$ and
$\gamma=\arg{(q^*\epsilon^{\prime})}$. Best fit value of solar angle
$\theta_\odot\approx 34^\circ$, $\tan 2\theta _{\odot}\approx 2.5$. It
demands that $|\delta|\cos\frac{\rho}{2}$ and
$2\Delta-\Delta'-|\delta'|\cos\gamma$ have to be nearly same order in
magnitude. According to the Eq.\ (\ref{predic}), in order to generate
solar mass difference successfully along with the solar angle those
combination of parameters should have following expression
\begin{eqnarray}
|\delta|\cos\frac{\rho}{2}\approx\frac{\Delta
  m_{21}^2}{4\sqrt{2}m_0^2}\sqrt{\frac{\tan^22\theta_{12}}{1+\tan^22\theta_{12}}}\nonumber\\
2\Delta-\Delta'-|\delta'|\cos\gamma\approx\frac{\Delta
  m_{21}^2}{2m_0^2}\sqrt{\frac{1}{1+\tan^22\theta_{12}}}.
\label{paramsol}
\end{eqnarray}
Again to get atmospheric
mass difference in right range we should take
\begin{eqnarray}
|\delta'|\cos\gamma\approx\Delta m_{32}^2/4m_0^2.
\label{paramatm}
\end{eqnarray}
We can conclude from Eq.\ (\ref{paramsol},\ref{paramatm}) that
smallness of $\tilde{L}$ symmetry breaking parameters are required for
successful generation of light neutrino mass differences and mixing
angles. From Eq.\ (\ref{paramsol},\ref{paramatm}) we can say that
$|\delta'|\cos\gamma$ is larger in value compare to
$|\delta|\cos\frac{\rho}{2}$ and
$2\Delta-\Delta'-|\delta'|\cos\gamma$. So, three terms in
$2\Delta-\Delta'-|\delta'|\cos\gamma$ are required to cancel among
themselves to generate $\Delta m_{21}^2$ value correctly. So far we
didn't tell about neutrinoless double beta decay. The relevant
quantity is
\begin{eqnarray}
(m_\nu)_{ee}\approx
-m_0\exp^{-i\theta_p}.
\label{dbeta}
\end{eqnarray}
So, from the neutrinoless double beta decay bound Eq.\ (\ref{nldb}) we
can say that $m_0\le {\rm 0.89 eV}$. If $m_0$ is larger side of WMAP
bound smallness of symmetry violating parameters will be confirmed. As
larger value of $m_0$ also confirms quasi-degenerate mass spectrum, so
we can say that the softness of $\tilde L$ symmetry violation and
degeneracy of light neutrino mass spectrum are corelated. We have
completed our discussion about light neutrino phenomenology except low
energy CP violation in the leptonic sector. After successful
predictions of light neutrino phenomenology let us see the
leptogenesis part after $\tilde L$ symmetry violation and also let us
observe how much value $m_0$ will require for generation of right
sized baryon asymmetry.
\subsection{Leptogenesis}
\label{abl}
We have already shown the $\tilde L$ symmetry violating $M_R$ in Eq.\ 
(\ref{sbmdmr}). Our primary goal is to diagonalize $M_R$ which is in
same form of $m_\nu$ in Eq.\ (\ref{mnu}). In order to do this we first
diagonalize  $H'=M_RM_R^\dagger$. 
\begin{eqnarray}
M_R=\left(\begin{array}{ccc} p & \epsilon & \epsilon \\ \epsilon & \epsilon' & q\\ \epsilon & q & \epsilon'
  \end{array}\right)\equiv \left(\begin{array}{ccc} A' & B' & B'\\ B' & C' & D'\\ B' & D' & C'
  \end{array}\right)\nonumber\\
H'=M_RM_R^\dagger=\left(\begin{array}{ccc} P' & Q'^* &  Q'^* \\ Q' & R' & S'
    \\ Q' & S' & R'\end{array}\right)
\end{eqnarray}
where
\begin{eqnarray}
P' = \left|A'\right|^2 + 2\left|B'\right|^2 &\qquad& Q'
=A'^*B'+B'^*(C'+D')\nonumber\\
R'=\left|B'\right|^2 + \left|C'\right|^2 + \left|D'\right|^2 &\qquad&
S'=\left|B'\right|^2 + D'C'^* + C'D'^*
\end{eqnarray}
The eigenvalues of $H'$ are in similar form of eigenvalues of $H$ in
Eq.\ (\ref{eigen}), 
\begin{eqnarray}
M^2_{2,1}=\frac{P'+R'+S'\pm\sqrt{(R'+S'-P')^2+
8\left|Q'\right|^2}}{2}\qquad M_3^2=R'-S'
\label{rhmas}
\end{eqnarray}
and the diagonalizing matrix $V$  of $H'$ is in same form of $U$ in Eq.\
(\ref{mixmat1}), 
\begin{eqnarray}
V=\left(\begin{array}{ccc} J' & \frac{\sqrt{2}Q^{'*}J'}{M_2^2-P'}
    & 0 \\
-\frac{Q'J'}{M_2^2-P'} & \frac{J'}{\sqrt{2}} & -\frac{1}{\sqrt{2}} \\
-\frac{Q'J'}{M_2^2-P'} & \frac{J'}{\sqrt{2}} & \frac{1}{\sqrt{2}}\end{array}\right)\quad
{\rm where }\quad J'=\sqrt{\frac{M_2^2-P'}{M_2^2-M_1^2}}.
\label{mixmat2}
\end{eqnarray}
We define a quantity $r$ which is $\tan 2\theta'_{12}$ from $V$
matrix. It has similar expression like $\tan 2\theta_{12}$ of Eq.\
(\ref{mixpat}). So,
\begin{eqnarray}
r=\tan 2\theta'_{12}=\frac{2\sqrt{2}|Q'|}{R'+S'-P'}\approx \frac{2\sqrt{2}|\delta|\cos\frac{\rho}{2}}{\Delta'+|\delta'|\cos\gamma},
\label{r}
\end{eqnarray}
where we have used
\begin{eqnarray}
&&R'+S'-P'\approx 2|p|^2(\Delta'+|\delta'|\cos\gamma)\nonumber\\
&&Q'\approx
2|p|^2|\delta|\cos{\frac{\rho}{2}}\exp^{i\theta_Q'}
\label{Q'}
\end{eqnarray}
with $2\theta_Q'=\arg{(p*q)}=-2\theta_Q$.
This $V$ can diagonalize $M_R$ but the entries may not be real
positive. So we can write
\begin{eqnarray}
V^\dagger M_RV^*={\rm diag}(M_1\exp^{i2\theta_1'},
M_2\exp^{i2\theta_2'}, M_3\exp^{i2\theta_3'}).
\label{mrd1}
\end{eqnarray}
We can evaluate the phases from  Eq.\ (\ref{mrd1}) and the obtained
results are
\begin{eqnarray}
&&\exp^{i2\theta_1'}=\frac{1}{(M_2^2-M_1^2)(M_2^2-P')M_1}
\left\{A'(M_2^2-P')^2-4Q^{'*}B'(M_2^2-P')+2Q^{'*2}(C'+D')\right\}\nonumber\\
&&\exp^{i2\theta_2'}=\frac{1}{(M_2^2-M_1^2)(M_2^2-P')M_2}\left\{2Q'^2A'+4Q'B'(M_2^2-P')+(C'+D')(M_2^2-P')^2\right\}\nonumber\\
&&\exp^{i2\theta_3'}=\frac{1}{M_3}(C'-D').
\label{majp2}
\end{eqnarray}
This phases can be absorbed to $V$. The redefined $V$ is
\begin{eqnarray}
V'=VV_P \qquad {\rm where} \qquad V_P={\rm  diag}(\exp^{i\theta_1'},\quad \exp^{i\theta_2'},\quad\exp^{i\theta_3'}).
\label{vp}
\end{eqnarray}
With this $V'$ and $m_D$ of Eq.\ (\ref{sbmdmr}) we can construct
$m_D'\equiv m_DV'$. The relevant matrix $h$ for leptogenesis takes the form
\begin{eqnarray}
h=m_D'^\dagger m_D'=\left(\begin{array}{ccc} 
J'^2 \left\{a^2+\frac{2b^2|Q|^2}{(M_2^2-P')^2}\right\} &
  \frac{\sqrt{2}J'^2Q'(a^2-b^2)}{M_2^2-P'}\exp^{i(\theta_1'-\theta_2')} & 0 \\
\frac{\sqrt{2}J'^2Q'^*(a^2-b^2)}{M_2^2-P'}\exp^{i(\theta_2'-\theta_1')}
&  J'^2\left\{b^2+\frac{2a^2|Q|^2}{(M_2^2-P')^2}\right\} & 0\\
0 & 0 & b^2 \end{array}\right).
\label{hmat1}
\end{eqnarray}
$h_{13}=h_{23}=0$  ensure that $N_3$ decay asymmetry
$\varepsilon_3=0$. We only now concentrate remaining two right handed
neutrinos $N_{1,2}$. The important quantity for generating decay
asymmetry parameters $\varepsilon_{1,2}$ is
\begin{eqnarray}
{\rm Im}(h_{12}^2)=-{\rm Im}(h_{21}^2)=\frac{2(a^2-b^2)^2}{(M_2^2-M_1^2)^2}{\rm Im}(Q'^2\exp^{i2(\theta_1'-\theta_2')}),
\label{h12}
\end{eqnarray}
where
\begin{eqnarray}
&&\exp^{i2(\theta_1'-\theta_2')}\approx\frac{M_2}{M_1}\left[1-\frac{1}{\sqrt{1+r^2}}\left(\Delta'+|\delta'|\exp^{i\gamma}\right)\right]\exp^{-i2\theta_Q'}.
\label{t1t2}
\end{eqnarray}
Now using Eq.\ (\ref{rhmas})
let us write down the important ratio of the masses of right handed
neutrinos,
\begin{eqnarray}
x_{12}=\frac{M_2^2}{M_1^2}\approx1+2(\Delta'+|\delta'|\cos\gamma)\sqrt{1+r^2}.
\label{massrat}
\end{eqnarray}
Now using Eq.\ (\ref{rhmas}), Eq.\ (\ref{r}), Eq.\ (\ref{Q'}) and Eq.\
(\ref{t1t2}) to Eq.\ (\ref{h12}) we obtain
\begin{eqnarray}
{\rm Im}(h_{12}^2)=-{\rm Im}(h_{21}^2)\approx
-(a^2-b^2)^2\frac{r^2}{4(1+r^2)^{3/2}}\times |\delta'|\sin\gamma
\label{h121}
\end{eqnarray}
Due to the fact of smallness of $1-x_{12}$ and $1-x_{21}$ appreciable
enhancement of $\varepsilon_{1,2}$ for self energy part
$\varepsilon^S_{1,2}$ in Eq.\ (\ref{self}) is possible when resonant
condition
\begin{eqnarray}
M_2-M_1\approx\Gamma_{1,2}/2
\label{reson}
\end{eqnarray}
is satisfied which is equivalent to the equality of the denominators
of self energy term $\varepsilon_{1,2}^S$ in Eq.\ (\ref{self}). We
can rewrite this condition as
\begin{eqnarray}
&&|1-x_{12}|\approx\frac{h_{22}}{4\pi v^2}\nonumber\\
&&\Longrightarrow 2(\Delta'+|\delta'|\cos\gamma)\sqrt{1+r^2}\approx\frac{a^2}{4\pi v^2}
\label{reson1}
\end{eqnarray}
where we have used the Eq.\ (\ref{massrat}). From the relations in Eq.\ 
(\ref{ddpm0}) and Eq.\ (\ref{reson1}) we can find heavy neutrino scale
$M$ in terms of $m_0$ and $\Delta'$, $|\delta'|\cos\gamma$ and $r$,
\begin{eqnarray}
M\equiv|p|=\frac{a^2}{m_0}\approx \frac {8\pi v^2(\Delta'+|\delta'|\cos\gamma)\sqrt{1+r^2}}{m_0}.
\label{hevscale}
\end{eqnarray}
At resonance vertex part of $\varepsilon_i$ is negligible. Considering
only self energy part Eq.\ (\ref{self}) with resonant condition of
Eq.\ (\ref{reson1}) we have 
\begin{eqnarray}
\varepsilon_1\approx\varepsilon_1^S=\frac{1}{4\pi v^2h_{11}}{\rm Im}(h_{12}^2)\times\frac{1}{2(1-x_{12})}=-\frac{{\rm Im}(h_{12}^2)}{2h_{11}h_{22}}.
\label{epsi1}
\end{eqnarray}
As $h_{11}\approx h_{22}$, $1-x_{12}\approx -(1-x_{21})$ and ${\rm
  Im}(h_{12}^2)=-{\rm Im}(h_{21}^2)$, so
$\varepsilon_1\approx\varepsilon_2$. Using Eq.\ (\ref{h121}) and $h_{11}\approx h_{22}\approx a^2$ to Eq.\ (\ref{epsi1}) we have
\begin{eqnarray}
\varepsilon=\varepsilon_1\approx\varepsilon_2&\approx&\frac{(a^2-b^2)^2}{a^4}\times\frac{r^2}{8(1+r^2)^{3/2}}\times |\delta'|\sin\gamma\nonumber\\
&\approx&4\Delta^2\times\frac{r^2}{8(1+r^2)^{3/2}}\times |\delta'|\sin\gamma
\label{epsi2}
\end{eqnarray}
where we have used $b=a(1+\Delta)$ from Eq.\ (\ref{assum}). One point
we should mention that although $h_{13}=h_{23}=0$, $\Delta_{\rm CP}$ of
Eq.\ (\ref{dcp}) is non zero after $\tilde{L}$ violation as $M_2\ne
M_1$ and $Im(h_{12}^2)\ne 0$ .

CP asymmetry parameters $\varepsilon_i$ are related
to the leptonic asymmetry parameters through $Y_L$ as
\cite{Pilaftsis1,Nielsen:2001fy,Pilaftsis:2003gt}
\begin{eqnarray}
Y_L\equiv\frac{n_L-{\bar n}_L}{s}=\sum_i^3\frac{\varepsilon_i\kappa_i}{g_{*i}}
\label{leptasym}
\end{eqnarray}
where $n_L$ is the lepton number density, ${\bar n}_L$ is the
anti-lepton number density, $s$ is the entropy density, $\kappa_i$ is
the dilution factor for the CP asymmetry $\varepsilon_i$ and $g_{*i}$
is the effective number of degrees of freedom \cite{Roos:1994fz} at
temperature $T=M_i$. Value of $g_{*i}$ in the SM with three right
handed Majorana neutrinos is $112$. The baryon asymmetry $Y_B$
produced through the sphaleron transmutation of $Y_L$ , while the
quantum number $B-L$ remains conserved, is given by
\cite{Harvey:1990qw}
\begin{eqnarray}
Y_B=\frac{\omega}{\omega-1}Y_L \qquad {\rm with} \qquad \omega =\frac{8N_F+4N_H}{22N_F+13N_H},
\label{barasym}
\end{eqnarray}
where $N_F$ is the number of fermion families and $N_H$ is the number
of Higgs doublets. The quantity $\omega=28/79$ in Eq.\ (\ref{barasym})
for SM. Now we introduce the relation between $Y_B$ and $\eta_B$,
where $\eta_B$ is the baryon number density over photon number density
$n_\gamma$. The relation is \cite{Barger}
\begin{eqnarray}
\eta_B=\left.\frac{s}{n_\gamma}\right|_0Y_B=7.0394Y_B,
\label{yetar}
\end{eqnarray}
where the zero indicates present time. Now using the relations in Eqs.\
(\ref{leptasym},\ref{barasym}, \ref{yetar}), $\omega=28/79$ and $g_{*i}=112$ we have
\begin{eqnarray}
\eta_B=-3.451\times 10^{-2}\sum_i\varepsilon_i\kappa_i.
\label{etab}
\end{eqnarray}
This dilution factor $\kappa_i$ approximately given by \cite{Nielsen:2001fy,kolb,Giudice:2003jh}
\begin{eqnarray}
\kappa_i\simeq\frac{0.3}{K_i(\ln K_i)^{3/5}} \qquad {\rm with}\qquad  K_i=\frac{\Gamma_i}{H_i},
\label{kppa}
\end{eqnarray}
where $\Gamma_i$ is the decay width of $N_i$ and $H_i$ is Hubble
constant at $T=M_i$. Their expressions are
\begin{eqnarray}
\Gamma_i=\frac{h_{ii}M_i}{4\pi v^2}\qquad {\rm and} \qquad H_i=1.66\sqrt{g_{*i}}\frac{M_i^2}{M_P},
\label{gh}
\end{eqnarray}
where $v=246$GeV and $M_P=1.22\times 10^{19}$GeV. Thus we have 
\begin{eqnarray}
K_i=\frac{M_Ph_{ii}}{1.66\times 4\pi\sqrt{g_{*i}}v^2M_i}.
\label{K}
\end{eqnarray}
%
Now the quantities $K_1$ and $K_2$ from Eq.\ (\ref{K}) takes the form
\begin{eqnarray}
K=K_1\approx K_2\approx 4.5\times
10^{-3}\times\frac{M_Pm_0}{v^2}\approx 913.2\times \frac{m_0}{1{\rm eV}}
\label{K1}
\end{eqnarray}
where $h_{11}/M_1\approx h_{22}/M_2\approx a^2/|p|=m_0$. With this we have
\begin{eqnarray}
\kappa= \kappa_1\approx\kappa_2\approx \frac{0.3}{K(\ln
  K)^{3/5}}\approx 3.2\times 10^{-4}\times\frac{1{\rm eV}}{m_0(\ln
  K)^{3/5}}.
\label{kapa1}
\end{eqnarray}
Using  $\varepsilon$ from Eq.\
(\ref{epsi2}) and $\kappa$ of Eq.\ 
(\ref{kapa1}) to the expression of baryon asymmetry Eq.\ (\ref{etab}),
we have
\begin{eqnarray}
\eta_B &\approx& -3.451\times 10^{-2}\times 2\times\varepsilon\kappa\nonumber\\
&\approx& -6.9\times 10^{-2}\times 3.2\times 10^{-4}\times\frac{1{\rm eV}}{m_0(\ln
  K)^{3/5}}\times 4\Delta^2\times\frac{r^2}{8(1+r^2)^{3/2}}\times
|\delta'|\sin\gamma\nonumber\\
 &\approx& -2.3\times 10^{-5}\times\frac{1{\rm eV}}{m_0}\times\left(\frac{1}{\ln
   K}\right)^{3/5}\times
 4\Delta^2\times\frac{r^2}{8(1+r^2)^{3/2}}\times \frac{\Delta m_{32}^2}{4m_0^2}\tan\gamma
\label{etab1}
\end{eqnarray}
where we have used the expression of $\delta'$ from Eq.\ (\ref{paramatm}).
Now for the numerical estimation let us consider two cases. 

{\large Case}-I: Seeing Eq.\ (\ref{paramsol}) we first consider that $2\Delta\approx \Delta'+|\delta'|\cos\gamma\approx 
\frac{\Delta  m_{21}^2}{2m_0^2\sqrt{1+\tan^22\theta_{12}}}$ which give
$r\approx\tan 2\theta_{12}$ from Eq.\ (\ref{r}). With this we have,
\begin{eqnarray}
\eta_B&\approx&-2.3\times 10^{-5}\times\frac{1{\rm eV}}{m_0}\times\left(\frac{1}{\ln
   K}\right)^{3/5}\times\left(\frac{\Delta
    m_{21}^2}{2m_0^2}\right)^2\times\frac{\tan^22\theta_{12}}{8(1+\tan^22\theta_{12})^{5/2}}\times \frac{\Delta m_{32}^2}{4m_0^2}\tan\gamma\nonumber\\
&\approx& -1.04\times10^{-19}\times\left(\frac{1{\rm eV}}{m_0}\right)^7\times\left(\frac{1}{\ln
   K}\right)^{3/5}\tan\gamma
\label{case1}
\end{eqnarray}
To get $\eta_B\approx 6.0\times 10^{-10}$ we have to take
$m_0\approx\sqrt{\Delta m_{\rm atm}^2}=0.045$eV for $\gamma=-67^\circ$. We have the right handed neutrino
mass scale $M\approx 3.0\times 10^{13}$GeV from Eq.\ (\ref{hevscale}). In this case we have $\delta\approx
6.5\times10^{-3}$ from Eq.\ (\ref{paramsol}) for small $\rho$ and
$\delta'=0.63$ from Eq.\ (\ref{paramatm}). 

{\large Case}-II: Again seeing Eq.\ (\ref{paramsol}) and Eq.\ (\ref{paramatm})
we also can consider that $2\Delta\approx
\Delta'+|\delta'|\cos\gamma\approx \frac{\Delta m_{32}^2}{4m_0^2}$
which give $r\approx\frac{2\Delta m_{21}^2}{\Delta
  m_{32}^2}\times\sqrt{\frac{\tan^22\theta_{12}}{1+\tan^22\theta_{12}}}
$ from Eq.\ (\ref{r}). With this we have,
\begin{eqnarray}
\eta_B&\approx&-2.3\times 10^{-5}\times\frac{1{\rm eV}}{m_0}
\times\left(\frac{1}{\ln
  K}\right)^{3/5}\times\left(\frac{\Delta
    m_{32}^2}{4m_0^2}\right)^2\times\left(\frac{2\Delta
    m_{21}^2}{\Delta m_{32}^2}\right)^2\nonumber\\
&&\times\frac{\tan^22\theta_{12}}{8(1+
\tan^22\theta_{12})}\times \frac{\Delta m_{32}^2}{4m_0^2}\tan\gamma\nonumber\\
&\approx& -2.0\times10^{-18}
\times\left(\frac{1{\rm eV}}{m_0}\right)^7\times\left(\frac{1}{\ln
 K}\right)^{3/5}\tan\gamma.
\label{case2}
\end{eqnarray}
For this case to obtain $\eta_B\approx 6.0\times 10^{-10}$ we have to
take $m_0\approx 0.067$eV for $\gamma=-67^\circ$. In this case we have
the right handed neutrino mass scale $M\approx 1.7\times 10^{14}$ from
Eq.\ (\ref{hevscale}). Here the $\tilde{L}$ symmetry violating
parameters will be $\delta\approx 2.9\times10^{-3}$ and
$\delta'\approx 0.28$ from Eq.\ (\ref{paramsol}) and Eq.\ 
(\ref{paramatm}) respectively. Again here we also have right sized
baryon asymmetry with right sign compatible with the light neutrino
phenomenology under soft breaking of $\tilde{L}$ symmetry. In Case-I
fine cancellation between $\Delta'$ and $|\delta'|\cos\gamma$ is
required. For Case-II fine cancellation is required between $2\Delta$
and $\Delta'+|\delta'|\cos\gamma$. In both cases we keep $\gamma$ in
the fourth quadrant ($\gamma=-67^\circ$) to get $\tan\gamma$ negative
which generates positive sign of baryon asymmetry. From Eq.\ 
(\ref{paramatm}) this corresponds to the fact that $\Delta m_{32}^2>0$
because $\cos\gamma$ is positive there. Now if we keep $\gamma$
($\gamma=113^\circ$ for both cases) in the 2nd quadrant we can have the
same results for baryon asymmetry.  But, for this $\gamma$ we can have
$\Delta m_{32}^2<0$ from Eq.\ (\ref{paramatm}). So, normal and
inverted ordering of quasi degenerate mass spectrum of light neutrinos
can be obtained under proper choice of $\gamma$. From the discussion
of the above two cases we can say that if $M\le 10^{12}$
\cite{Grimus:2003sf,Pilaftsis1,Nielsen:2001fy} this bound is not so
strict we can have proper sized baryon asymmetry with right sign
maintaining light neutrino phenomenology in tact under soft breaking
of $\tilde{L}$ symmetry.

In SM framework RG evolution effects on mixing angles and masses are
usually small because neutrino Yukawa couplings are small. In our case
neutrino Yukawa couplings are not small and the netrino mass spectrum is
quasidegenerate kind. So, RG effect may be appreciable on light
neutrino mass and mixing pattern
\cite{Rodejohann:2005ru,Antusch:2005gp}. We have checked that in
Case-I mixing pattern of light neutrino is stable under RG running.
But, in Case-II solar angle has an appreciable effect from GUT scale
to see-saw scale keeping other angles unaltered. We have also seen
that RG evolution of $m_0^2$, $\Delta m_{21}^2$ and $\Delta m_{32}^2$
are dominated by $\bar{\alpha}_{SM}$ \cite{Antusch:2005gp}. The proper
choice of GUT scale value of $m_0^2$, $\Delta m_{21}^2$ and $\Delta
m_{32}^2$ in both cases and solar angle in Case-II can match the low
energy data of light neutrino. We also have observed that
$\frac{m_0^2}{\Delta m_{21}^2}$, $\frac{m_0^2}{\Delta m_{32}^2}$,
$\frac{\Delta m_{21}^2}{\Delta m_{32}^2}$ are almost independent of
scale. It ensures that RG evolution has a little effect on our results
of baryon asymmetry in both cases.
\section{Conclusion}
\label{conclu} 
We have studied light neutrino phenomenology for the softly broken
$\tilde{L}$ symmetry keeping $S_2^{\mu\tau}$ symmetry which is an
accidental symmetry of light and heavy Majorana neutrino mass terms
for the exact $\tilde{L}$ symmetry. We restrict ourselves to type-I
See-Saw mechanism for generation of light neutrino mass term. Under
the assumption of $\tilde{L}$ symmetry preserving non zero parameters
in the neutrino sector are nearly same, we can have quasi degenerate
mass spectrum for both light and heavy Majorana neutrino. We have seen
that softness of the $\tilde{L}$ symmetry violation depends on the
degeneracy of the light neutrino mass spectrum. More degeneracy
implies more soft $\tilde{L}$ violation. Common light neutrino mass
scale can be as large as the upper bound from WMAP data. Soft breaking
of $\tilde{L}$ symmetry have an ability to generate light neutrino
phenomenology. We are in $\theta_{13}=0$ and
$\theta_{23}=\frac{\pi}{4}$ case. In \cite{Choubey:2004hn} authors
consider radiative correction in addition to soft breaking of
$\tilde{L}$ symmetry. They need large $\tan\beta$ for generation
appreciable radiative correction from SUSY.  They predicted non-zero
$\theta_{13}$ and hierarchy of solar and atmospheric mass difference.
We restrict ourselves to SM framework with three singlet right handed
neutrinos. We have found that the exact $\tilde{L}$ symmetry forbids
leptogenesis. We have studied whether it is possible or not to
generate right sized baryon asymmetry after breaking of $\tilde{L}$
symmetry. The quasi-degenerate mass spectrum of right handed neutrinos
opens an option for resonant kind of leptogenesis. We have found that
the degeneracy of right handed neutrino mass spectrum is restricted
through light neutrino data. We have found that common light neutrino
mass scale $m_0$ have to be of the order of $\sqrt{\Delta m^2_{\rm
    atm}}$ for generation of WMAP value of baryon asymmetry for one
case. The resonant condition generates the scale of heavy Majorana
neutrino mass which is nearly $10^{13}$ GeV in that case. Little
larger value of $m_0$ and $M$ are required for another case to
generate baryon asymmetry. If the bound $M\le 10^{12}$GeV as pointed
out in \cite{Grimus:2003sf,Pilaftsis1,Nielsen:2001fy} is confirmed
present model could not predict baryon asymmetry. Then the baryon
asymmetry have to have different origin other than the heavy Majorana
neutrino decay generated leptogenesis and its transmutation to baryon
asymmetry through sphaleron. Again we have seen that both normal and
inverted types of ordering of quasidegenearte mass spectrum of light
neutrinos are compatible with baryon asymmetry data.  We have checked
that solar angle in Case-II and $m_0^2$, $\Delta m_{21}^2$ and $\Delta
m_{32}^2$ masses in both cases have appreciable RG effects.  Proper
choice of GUT scale value of those quantities can match the low energy
data of light neutrino. We have also seen that RG evolution has a
little effect on the predictions of baryon asymmetry.  Another point
we should mention that the $S_2^{\mu\tau}$ is broken in the charged
lepton mass term. To break $S_2^{\mu\tau}$ naturally more two scalar
doublets $\Phi'$ and $\Phi''$ can be adapted as in
\cite{Grimus:2003sf}. In this \cite{Grimus:2003sf} article authors
considered that $\Phi$ couples to $e_R$, $\Phi'$ couples to $\mu_R$
and $\Phi''$ couples to $\tau_R$.  Under $S_2^{\mu\tau}$
$\mu\leftrightarrow\tau$, the $\Phi'$ remains invariant while $\Phi''$
changes sign; this leads to $m_\mu\ne m_\tau$. Incorporation of the
two scalar does not affect our light neutrino phenomenology. The
effective degrees of freedom $g*$ changes $112$ to $120$ and $\omega$
changes $28/79$ to $12/35$ which cannot change appreciably of our
predictions in baryon asymmetry.
\paragraph*{Acknowledgments}
The author acknowledges helpful discussions with Palash B. Pal. He
also thanks him for reading the manuscript and suggesting
improvements. The author also thanks Michael Schimdt for his valuable
comments about the RG effect.

\end{document}